\documentclass[pre,aps,nofootinbib,twocolumn]{revtex4}
\usepackage{amssymb}
\usepackage{graphicx}
\usepackage{amsmath}
\usepackage{natbib}

\input{tcilatex}
\begin{document}

\title{Comment on paper arXiv: 17070526v1 "Electronic structure of FeSe
monolayer superconductors: shallow bands and correlations" by Sadovskii group%
}
\author{ $^{1,2}$ Miodrag L. Kuli\'{c}}

\address{$^{1}$Institute for Theoretical Physics, Goethe-University D-60438
Frankfurt am Main, Germany \\
$^{2}$Institute of Physics, Pregrevica 118, 11080 Belgrade,
Serbia}

\begin{abstract}
We comment two incorrect statements given in \cite{Sadovskii}. (\textbf{A})
- In order to show that the electron-phonon interaction (EPI)\ is very small
and irrelevant for high T$_{c}$ superconductivity in 1UC FeSe/SrTiO$_{3}$
system, the authors of \  \cite{Sadovskii} use an EPI\ coupling constant ($%
\lambda _{Sad}$ ) which does not enter in any theory of superconductivity.
So, their conclusion on the smallness of the EPI in 1UC FeSe/SrTiO$_{3}$
\textit{is} \textit{incorrect}. Accordingly, their coupling constant $%
\lambda _{Sad}$ has also nothing to do with the EPI coupling with the
forward scattering peak (EPI-FSP), which is proposed recently in order to
explain high T$_{c}$ in 1UC FeSe/SrTiO$_{3}$. (\textbf{B}) - In \cite%
{Sadovskii} it is claimed that the experimentally resolved ARPES replica
bands can be explained by the LDA+DMFT method of Ref. \cite{Sadovskii}. We
show that this statement is also incorrect, i.e. the LDA+DMFT method is
unable to explain the replica bands.
\end{abstract}

\date{\today }
\maketitle

\address{$^{1}$Institute for Theoretical Physics, Goethe-University D-60438
Frankfurt am Main, Germany \\
$^{2}$Institute of Physics, Pregrevica 118, 11080 Belgrade,
Serbia}

\section{Introduction}

The recent discovery of high temperature superconductivity(SC) in the\textit{%
\ one unit-cell film} of the iron-selenide $FeSe$ grown on the $SrTiO_{3}$
substrate - called $1UC$ $FeSe/SrTiO_{3}$, with $T_{c}\sim 100$ $K$ , as
well as grown on the rutile $TiO_{2}$ (100) substrate with $T_{c}\sim 65$ $K$
\cite{Wang}, has provoked an intensive debate on the origin of SC in this
system. Additionally, ARPES spectra give strong evidence for the existence
of replica bands with the same shape as the main electronic band responsible
for SC. The replica bands are shifted by $\sim 100$ $meV$. In that respect
in \cite{Lee-Interfacial}-\cite{Johnston1-2} is proposed that these
experimental facts can be consistently explained by the theory of the
electron-phonon interaction with the forward scattering peak - the EPI-FSP
theory. The latter theory is proposed in \cite{Kulic-Zeyher1-2}, while its
extreme case with the delta-like peak is elaborated in \cite{Dan-Dol-Kul-Oud}%
. In \cite{Kulic-Dolgov-NJP} some important issues were elaborated and
cleared up. Additionally, the range of microscopic parameters relevant for $%
1UC$ $FeSe/SrTiO_{3}$ is estimated. The basic assumption of \ the \textit{%
EPI-FSP theory} is that the transverse \textit{oxygen optical phonon }due to
the $TiO_{2}$ layer with the frequency $\Omega _{O}\sim 90$ $meV$ is the
main pairing glue and that the corresponding EPI pairing potential is peaked
at small transfer momenta $q\approx 0$, i.e. $g(q)=g_{0}\exp \{-q/q_{c}\}$
with $q_{c}\ll k_{F}$. The important predictions of the EPI-FSP theory are: $%
(A)$ the SC critical temperature $T_{c}$ and the gap $\Delta $ are\textit{%
linear functions on the pairing potential}, i.e. $T_{c}\sim V_{FSP}/4$, $%
\Delta =2T_{c}$, $V_{FSP}\sim (q_{c}/G)^{2}(2g_{0}^{2}/\Omega _{O})$, $G=\pi
/a$ and $a$ is the lattice constant. In the derivation of these results it
is assumed that $q_{c}v_{F}<\pi T_{c}\ll \Omega _{O}$. Note, that $T_{c}$
and $\Delta $ \textit{do not depend on the oxygen mass} - since $V_{FSP}$ is
mass independent in leading order; $(B)$ in ARPES spectra \textit{there are
sharp replica bands }with the same shape as the main band and shifted by the
multiple of energy $\sim \Omega _{O}$. These results are contrary to the
standard \textit{isotropic Eliashberg theory} (\textit{ET}) where $T_{c}$ is
mass-dependent, i.e. $T_{c}^{ET}\sim M_{O}^{-1/2}$, and the replica bands
would be drastically deformed. At the same time the low-energy ($\omega \ll
\Omega _{O}$) slope of the self-energy $\Sigma (\omega )$ is mass-dependent
in the EPI-FSP theory, while in the ET it is not, i.e. $\Sigma _{FSP}(\omega
)\sim -\lambda _{m}\omega $ with $\lambda _{m}\sim \sim M_{O}^{1/2}$. \ This
means that \textit{the predictions of the EPI-FSP theory are very different
from the ET theory}. The ARPES results in $1UC$ $FeSe/SrTiO_{3}$ \cite%
{Lee-Interfacial} are compatible with $\lambda _{m}\sim 0.1-0.2$ as shown in
\cite{Johnston1-2} and \cite{Kulic-Dolgov-NJP}. Note, that in spite of the
fact that $\lambda _{m}$ is rather moderate, high $T_{c}$ is reached thanks
to the linear dependence, $T_{c,FSP}\sim V_{FSP}$, instead of the
exponential one in the ET theory, $T_{c,ET}\sim \exp \{-1/\lambda _{ET}\}$.

Recently, intensive efforts were done in order to discredit and disregard
the EPI mechanism of pairing and its origin of the replica bands in $1UC$ $%
FeSe/SrTiO_{3}$. These approaches are mainly based on the spin-fluctuations
interaction described by the extended Hubbard or phenomenological Heisenberg
models. For instance, the Sadovskii's group \cite{Sadovskii} claims to have
shown: $(A)$ that the EPI coupling is extremely small and irrelevant for
high $T_{c}$ and $(B)$ - the replica bands are due to strong correlations in
the LDA-DMFT approach. \textit{Let us show that both claims are incorrect}.

$(A)$ \textbf{Role of EPI on }$T_{c}$ - In order to show that the EPI is
irrelevant in $1UC$ $FeSe/SrTiO_{3}$, i.e. that $T_{c}$ due to EPI-FSP is
small, in \cite{Sadovskii} this problem is studied in the framework of the
ET theory (with the band energy $\xi _{\mathbf{p}}=\varepsilon _{\mathbf{p}%
}-\mu $ and with the Einstein phonon with the energy $\Omega _{O}$) by
calculating a \textit{quite inappropriate coupling constant} $\lambda _{Sad}$
($N$ is the number of unit cells)

\begin{equation}
\lambda _{Sad}=\frac{2}{N\Omega _{O}}\frac{\sum_{\mathbf{p},\mathbf{q}}\left
\vert g(\mathbf{q})^{2}\right \vert \delta (\xi _{\mathbf{p}})\delta (\xi _{%
\mathbf{p}+\mathbf{q}}-\Omega _{O})}{\sum_{p}\delta (\xi _{\mathbf{p}})}
\label{lambda-sad-1}
\end{equation}%
\begin{equation}
\lambda _{Sad}\sim \lambda _{0}\frac{\Omega _{O}}{\pi \varepsilon _{F}}\sqrt{%
\frac{q_{c}v_{F}}{\Omega _{O}}}\exp (-\frac{2\Omega _{O}}{q_{c}v_{F}})\sim
10^{-9}\lambda _{0},  \label{lambda-sad-2}
\end{equation}%
for experimental values ($\Omega _{O}/\pi \varepsilon _{F}$)$\sim 1$ and ($%
\Omega _{O}/q_{c}v_{F}$)$\sim 10$. If this analysis were correct it would
give an enormous small $T_{c,Sad}\sim \exp (-10^{9})$ $K$ in the ET theory.
However, \textit{the coupling} $\lambda _{sad}$\textit{\ never appears in
any theory of superconductivity}! Namely, in the ET theory by assuming that
the phonon line-width $\Gamma _{0}$ is much smaller than the phonon energy $%
\Omega _{O}$, i.e. $\Gamma _{0}\ll \Omega _{O}$, the critical temperature $%
T_{c}$ is determined by the coupling constant $\lambda _{ET}$ defined by

\begin{equation}
\lambda _{ET}=\frac{2}{N\Omega _{O}}\frac{\sum_{\mathbf{p},\mathbf{q}%
}\left \vert g(\mathbf{q})^{2}\right \vert \delta (\xi _{\mathbf{p}})\delta
(\xi _{\mathbf{p}+\mathbf{q}})}{\sum_{p}\delta (\xi _{\mathbf{p}})}\sim
\lambda _{0}\frac{q_{c}}{4\pi k_{F}},  \label{lambda-eli}
\end{equation}%
i.e.
\begin{equation}
\lambda _{ET}\gg \lambda _{Sad}\text{ }!  \label{lambda-eli-sad}
\end{equation}%
It is clear that the \textit{Eliashberg coupling is much larger} than the
one introduced and calculated by Sadovskii's group, i.e. $\lambda _{ET}\gg
\lambda _{Sad}$. It is physically clear why $\lambda _{Sad}$ cannot be
related to SC, since it describes real scattering of electrons on phonons
where one optical phonon is emitted (or absorbed). This is seen in $Eq.$(\ref%
{lambda-sad-1}) where $\lambda _{Sad}$ contains two delta functions which
describe conservation of energy in the scattering processes. On the other
side the Eliashberg coupling $\lambda _{ET}$ describes virtual excitation
and absorption of phonons by electrons, which are responsible for the mass
renormalization and superconductivity. In conclusion, in \cite{Sadovskii}
\textit{the EPI coupling constant is enormously underestimated by nine order
of magnitude due to using quite inappropriate EPI coupling constant}. We
point out, that in spite of the fact that $\lambda _{ET}\gg \lambda _{Sad}$
the ET approach would still give rather small $T_{c}$, since for an
optimistic estimation one has $\lambda _{ET}\sim 0.1$ and $T_{c,ET}$ is
rather small, i.e. ($T_{c,Sad}\ll $)$T_{c,ET}\sim \Omega _{0}\exp
(-1/\lambda )<0.01$ $K$. In that respect the\textit{\ }recently proposed%
\textit{\ EPI-FSP mechanism} of pairing in $1UC$ $FeSe/SrTiO_{3}$ \cite%
{Johnston1-2}, \cite{Kulic-Dolgov-NJP} \textit{is much more favorable} due
to its linear dependence of $T_{c}$ on the pairing potential, \ i.e. $%
T_{c,FSP}\sim (q_{c}/G)^{2}(2g_{0}^{2}/\Omega _{O})=(q_{c}/G)^{2}V_{FSP}^{0}$%
. It is matter of fine nature-tuning that in $1UC$ $FeSe/SrTiO_{3}$ the
reasonable value for $(q_{c}/G)\sim 0.1-0.2$ is realized and for $%
V_{FSP}^{0}\sim (0.5-1)$ $eV$ one has $T_{c}\sim 100$ $K$ \cite{Johnston1-2}%
, \cite{Kulic-Dolgov-NJP}. In reality the EPI-FSP pairing mechanism acts not
alone, since one should add a "residual" pairing which is responsible for SC
in the single $FeSe$ plane but with the electron-like Fermi surface at the
point $M$. This means that $T_{c,FSP}$ is a lower bound of $T_{c}$, i.e. $%
T_{c,FSP}<T_{c}$. \ We stress again, that the the EPI-FSP theory is already
partly confirmed by the perfect shape of the ARPES replica bands \cite%
{Lee-Interfacial} and by the mass-dependent of their energy shift with
respect to the main electronic band and by the linear dependence of $\Delta $
on the coupling strength \cite{Song}. Measurements of the oxygen
mass-independence of $T_{c}$ and $\Delta $, as well as of the
mass-dependence of the self-energy - predicted in  \cite{Kulic-Dolgov-NJP},
would be an important step in proving the relevance of the EPI-FSP pairing
mechanism in $1UC$ $FeSe/SrTiO_{3}$.

The above discussion shows that the claims done in \cite{Sadovskii} - on the
weakness of EPI in $1UC$ $FeSe/SrTiO_{3}$ are unfounded, since the analysis
in \cite{Sadovskii} is based on an inappropriate coupling constant and on an
inappropriate EPI theory. In that respect any eventual reply on our comment
of the point $(A)$ is superfluous.

\begin{figure}[hbtp]
\centerline{\includegraphics[clip,
width=1.0\columnwidth]{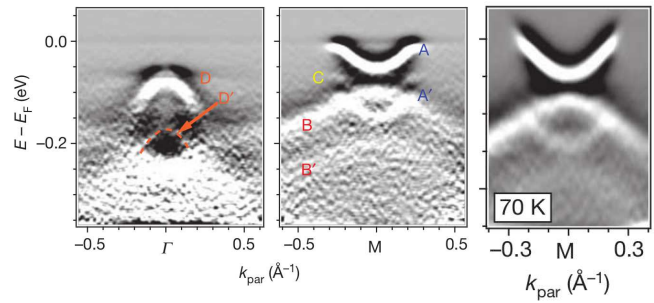}} \caption{ARPES bands in $1UC$
$FeSe/SrTiO_{3}$ - from \protect \cite{Lee-Interfacial}.
\textit{Left}: The hole band $D$ and its replica $D^{\prime }$ at
the $\Gamma$ point; \textit{Middle}: The electronic band $A$ and
its replica $A^{\prime }$ at the M point, shifted by $100$ $meV$
at $T<T_{c}$; \textit{Right}: The
electronic band $A$ and its replica $A^{\prime}$ at the M point, shifted by $100$ $meV$ at $%
T>T_{c}$.} \label{Fig1}
\end{figure}

$(B)$ \textbf{Sharp replica bands }- Important ARPES results related to the
energy spectra of $1UC$ $FeSe/SrTiO_{3}$ are reported in \cite%
{Lee-Interfacial} - see Fig.1: ($i$) two electronic (almost degenerate)
bands are clearly resolved with the Fermi surfaces centered at the point $M$
- these bands are labelled by $A$ with the band bottom energy $\sim 60$ $meV$
from the Fermi surface; ($ii$) the existence of the replicated bands $%
A^{\prime }$ with the same shape as $A$ but shifted downward by the energy $%
\sim 100$ $meV$ - this energy is of the order of the optical phonon energy $%
\Omega _{0}$; ($iii$) in the superconducting state both bands $A$ and $%
A^{\prime }$ show typical superconducting behavior with banding away from
the Fermi surface (at $k_{F}$) - see the backbending at $k_{F}$ in
Fig.1(Middle); ($iv$) in \cite{Song} it is found the \textit{isotope effect
in the energy shift} ($\sim M_{0}^{-1/2}$) of the band $A^{\prime }$ with
respect to $A$.

However, in \cite{Sadovskii} it is claimed that the replica band $A^{\prime
} $ can be explained (even semi-quantitatively) exclusively in the LDA-DMFT
approach, i.e. to be due to electronic correlations. They found that the
electronic band $A$ is formed by the $Fe-3d_{xz},3d_{yz}$ states while its
bottom energy is $\sim 100$ $meV$, i.e. larger than the experimental value,
while the band $A^{\prime }$ is due to the $Fe-3d_{xy}$ state. However, the
band $A^{\prime }$ does not have the shape and properties of the $A$ band -
see Fig.2. Namely, the ARPES replica band $A^{\prime }$ exists for $k<k_{F}$
and is reminescent of the vibron shake-offs in the photoemission of H$_{2}$
molecule \cite{Lee-Interfacial}, \cite{Turner}, while the LDA-DMFT (ARPES)
band $A^{\prime }$ goes up to the Fermi surface. In fact the band $A^{\prime
}$ from \cite{Sadovskii} is more reminescent of the second almost degenerate
band around the point $M$, as reported in \cite{Lee-Interfacial}.

\begin{figure}[tbph]
\centerline{\includegraphics[clip,
width=1.0\columnwidth]{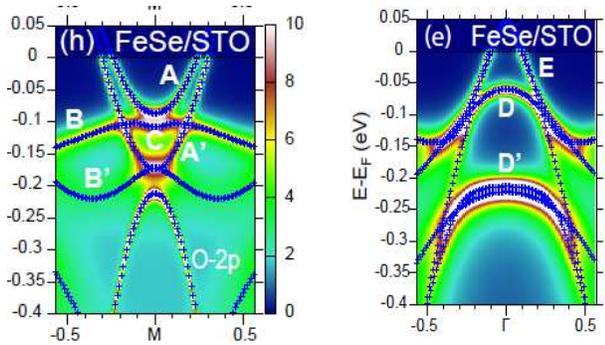}} \caption{The
LDA-DMFT electronic bands - from \protect \cite%
{Sadovskii}. \textit{Left}: The bands $A$ and $A^{\prime }$ at the
M point in the normal state; \textit{Right}; The hole bands $D$
and $D^{\prime }$ at the $\Gamma$ point as well as the artefact
band $E$.} \label{Fig2}
\end{figure}

Moreover, the LDA-DMFT brings in unpleasent artefact since it predicts an
hole-like Fermi surface near the $\Gamma $ point with the band-top at $\sim
50$ $meV$. Experimentally there is only a hole band $D$ at the $\Gamma $
point, which is approximately $40$ $meV$ \ below the Fermi surface and its
replica band $D^{\prime }$ shifted downward by the energy $\sim \Omega _{0}$%
. This means that \textit{the experimental results} ($i$)-($iv$) \textit{%
cannot be explained by the} LDA-DMFT \textit{approach}.

In conclusion, the band spectra with their replica bands and high $T_{c}$
superconductivity in $1UC$ $FeSe-SrTiO_{3}$ cannot be explained by the
LDA-DMFT approach proposed in \cite{Sadovskii}. Contrary to LDA-DMFT, the
theory based on the electron-phonon interaction with the forward scattering
peak (EPI-FSP) is able to explain some important experimental facts by
including the isotope effect in the shift of the replica bands.


\begin{thebibliography}{9}
\bibitem{Sadovskii} I. A. Nekrasov, N. S. Pavlov, M. V. Sadovskii, arXiv:
1707.052626v1 (2017)

\bibitem{Wang} Wang Quing-Yan, Li Zhi, Zhang Wen-Hao, Zhang Zuo-Cheng, Zhang
Jin-Song, Li Wei, Ding Hao, Ou Yun-Bo, Deng Peng, Ghang Kai, Wen Jing, Song
Can-Li, He Ke, Jia Jin-Feng, Ja Shuai-Hua, Wang Ya-Yu, Wang Li-Li, Chan Xi,
Ma Xu-Cun, Xue Qi-Kun, Chin. Phys. Lett. \textbf{29}, 037402 (2012)

\bibitem{Lee-Interfacial} J. J. Lee, F. T. Schmitt, R. G. Moore, S.
Johnston, Y. T. Cui, W. Li, M. Yi, Z. K. Liu, Y. Zhang, D. H. Lu, T. P.
Devereaux, D.-H. Lee, Z. X. Shen, Nature \textbf{515}, 245 (2014)

\bibitem{Johnston1-2} L. Rademaker, Y. Wang, T. Berlijn, S. Johnston, New J.
Phys. \textbf{18}, 022001 (2016); Y. Wang, K. Nakatsukasa, L. Rademaker, T.
Berlijn, S. Johnston, Supercond. Sci. Technol. \textbf{29}, 054009 (2016)

\bibitem{Kulic-Zeyher1-2} M. L. Kuli\'{c}, R. Zeyher, Phys. Rev. {\textbf{B }%
49}, 4395 (1994); R. Zeyher, M. L. Kuli\'{c}, Phys. Rev. {\textbf{B }53},
285 (1996); M. L. Kuli\'{c}, Phys. Reports {\textbf{3}38}, 1-264 (2000)

\bibitem{Dan-Dol-Kul-Oud} O. V. Danylenko, O. V. Dolgov, M. L. Kuli\'{c}, V.
Oudovenko, Europ. Phys. Jour. {\textbf{B}9} - Cond. Matter, 201 (1999)

\bibitem{Kulic-Dolgov-NJP} M. L. Kuli\'{c}, O. V. Dolgov, New J. Phys.
\textbf{19}, 013020 (2017)

\bibitem{Song} Q. Song, T. L. Yu, X. Lou, B. P. Xie, H. C. Xu, C. H. P. Wen,
Q. Yao, S. Y. Zhang, X. T. Zhu, J. D. Guo, R. Peng, D. L. Feng, arXiv:
1710.07057v1 (2017)

\bibitem{Turner} D. Turner, \textit{Molecular Photoelectron Spectroscopy}%
(Wiley, 1970)
\end{thebibliography}
\end{document}